\documentclass{article}
\usepackage{graphicx}
\usepackage{amsmath}
\usepackage{amsfonts}
\usepackage{amssymb}
\newtheorem{theorem}{Theorem}
\newtheorem{acknowledgement}[theorem]{Acknowledgement}

\begin{document}

\title{On Zeeman Topology in Kaluza-Klein and Gauge Theories}
\author{\textbf{Ivan Struchiner}\thanks{E-mail address: ivan@ime.unicamp.br} \textbf{
and M\'{a}rcio A. F. Rosa}\thanks{E-mail address: marcio@ime.unicamp.br}}
\maketitle
\begin{abstract}
E. C. Zeeman \cite{1} has criticized the fact that in all articles and books
until that moment (1967) the topology employed to work with the Minkowski
space was the Euclidean one. He has proposed a new topology, which was
generalized for more general space-times by Goebel \cite{2}. In the Zeeman and
Goebel topologies for the space-time, the unique continuous curves are
polygonals composed by time-like straight lines and geodesics respectively. In
his paper, Goebel proposes a topology for which the continuous curves are
polygonals composed by motions of charged particles. Here we obtain in a very
simple way a generalization of this topology, valid for any gauge fields, by
employing the projection theorem of Kaluza-Klein theories (page 144 of
Bleecker \cite{3}). This approach relates Zeeman topologies and Kaluza-Klein,
therefore Gauge Theories, what brings insights and points in the direction of
a completely geometric theory.
\end{abstract}

\section{Introduction}

In 1967, E. C. Zeeman \cite{1} criticized the topology usually employed to
word with the Minkowski space and proposed a substitute. These criticisms were
done by Cel'nik \cite{4} in the following year. In all books and articles
until then the topology employed for Minkowski space was the Euclidean
topology of the $4-$dimensional space. This was done in a straight way or
implicitly when operations involving limits were performed.

Despite the developements of Zeeman's proposal, done for Lorentzian manifolds
by Goebel (\cite{4},\cite{5}), Hawking-King-MacCarthy \cite{6}, Domiaty
(\cite{7},\cite{8},\cite{9}) and Malament \cite{10}, the situation has not
changed nowadays, almost all books and articles on Minkowski space and other
space-times have employed the Euclidean topology of the $4-$dimensional
Euclidean Space and its correspondent generalizations. It seems that for the
good idea of Zeeman it was not found yet a sounding application.

The topology proposed by Zeeman in his seminal article is the finest topology
that induces the $3-$dimensional Euclidean topology on every space axis and
the 1-dimensional Euclidean topology in every time axis. For this topology the
continous curves in the Minkowski space M are polygonals composed by straight
time-like paths and the group of homeomorphisms of M becomes the non
homogeneous Lorentz group increased by dilatations. The space-like axis is a
hypersurface with a time-like normal vector and a time-like axis is a straight
time-like line. All the proofs in Zeeman paper remain valid if instead of
Minkowski space time he had employed a $1+n$ dimensional linear
semi-Riemannian manifold with one time-like and $n>3$ space-like dimensions.
Then the group of homeomorphisms for the Zeeman topology would be the
generalization of the non homogeneous Lorentz group, that is the semidirect
product of the $n-$dimensional translations with the connected subgroup of
$O(1,n)$ increased by the dilatations. Some generalizations for a space with
$p$ space-like and $q$ time-like dimensions were studied for $p$ and $q$ not
equal in \cite{11}.

Goebel \cite{3} has generalized the Zeeman topology for more general
Lorentzian manifolds, where the continous curves would be geodesic polygonals,
that is, composed by timelike geodetic paths. It is well known that in the
Riemannian case \cite{4} we can recover the topology from the metric and
geodesics by defining a basis of open normal balls. What Goebel has done was
to construct from the Lorentzian metric the local neighborhoods of the Zeeman
topology, obtaining its generalization for a Lorentzian manifold. Also the
proofs of Goebel paper are all valid if we had as space-time a semi-Riemannian
manifold with one time-like and $n>3$ space-like dimensions. Goebel also
proposed to change the geodesics in his construction by the motion of
eletrically charged particles, obtaining a topology for which a continuous
curve is
$<$%
$<$%
a chain of finitely many connected world lines of freely falling charged test
particles%
$>$%
$>$%
(page 296 of \cite{3}).

Hawking, King and MacCarthy have proposed another topology \cite{6} for which
the continuous paths included all the possible time-like paths, there they
argue that this would be physically more interesting, since $\ll$even in general
relativity, particles need not move along geodesics since, for example, they
may be charged and an electromagnetic field may be present (and this applies
to special relativity also)$\gg$. But if we intend to have a completely geometric
theory in which all the fields acting on the particle are included in the
space-time structure, a particle has no more to do in an empty space-time than
follow geodesics, or straight time-like paths if we are in the Minkowski case.

On the other hand, Zeeman topology gives a polygonal composed of many straight
time-like paths and not only one. The interpretation given originally by
Zeeman \cite{1} as the...
$<$%
$<$%
path of a freely moving particle under a finite number of collisions%
$>$%
$>$%
... which was repeated many times in the literature can also be criticized in
the same way since, travelling in the empty space, a particle has no target to
hit. Analogous problem has some interpretation of Goebel's generalization,
where we have polygonals of geodesic paths.

But the situation is not so bad in the Goebel and Zeeman cases since it is
possible to differentiate along time-like straight lines and it is possible to
define a massive particle's path as a continuous and differentiable path,
avoiding the zig-zags. It is enough to discover the wheel, the axiomatic gain
is yet very big.

Here, following the idea of obtaining a completely geometric theory, that we
turn to Goebel's charged particles topology. We employ the projection theorem
for Kaluza-Klein theories (part 2) to present it in a very simple way that
works as well to general gauge fields (part 3). This approach relates Zeeman
topologies and Gauge Theories, bringing insights and suggesting interesting
conjectures and developements (part 4) and from the universality of the gauge
theories, it points in the direction of a completely geometric theory.

For this, in what follows we will not deal with polygonals of, but with
geodesics, any way the results, valid path by path, could be rewritten in
terms of polygonals.

\section{\textbf{The Projection Theorem in Gauge and Kaluza-Klein Theories}}

The Kaluza-Klein theories are the classical model for Gauge Theories. The
space-time of such theories is the total space $P$ of a fiber bundle
$\pi:P\rightarrow M$ (here we shall follow Bleecker, \cite{2}). As seen above
it is natural to define on $P$ a metric which turns it into a semi-Riemannian
manifold with signature $(1,n),\;n=3+dim\;G$ , where the number of
\ space-like dimensions is three plus the dimension of the gauge group $G$ of
the theory, which is also the structure group, diffeomorfic to the fiber of
the bundle (here we suppose the Lie group $G$ compact). The Minkowski space,
or any other relativistic space-time M is the base space of the fiber bundle.

Additionally we suppose we have defined a connection 1-form $\omega\in
\Omega^{1}\left(  P,\;Lie\;G\right)  $ on P with values in the Lie algebra of
$G,Lie\;G$. This connection gives rise to a covariant derivative $D^{\omega
}:\Omega^{q}\left(  P\text{,...}\right)  \rightarrow\Omega^{q+1}\left(
P,...\right)  $ and associated curvature $\Omega^{\omega}=D^{\omega}\omega
\in\Omega^{2}\left(  P,\;Lie\;G\right)  .$ The fiber bundle is locally trivial
, that is, there is a cover $U_{\alpha}$ for $M$ and associated local sections
$\sigma_{a}:U_{\alpha}\rightarrow P$, with $\pi\circ\sigma_{a}=id_{\alpha}$,
the identity on $U_{\alpha}$. Each local section is interpreted as a gauge
choice, and gives rise to a gauge potential $iA_{\alpha}=\sigma_{\alpha}%
^{\ast}\omega_{\alpha}\in\Omega^{1}\left(  U_{\alpha}\text{, }Lie\;G\right)  $
and to a gauge field $iF_{\alpha}=\sigma_{\alpha}^{\ast}\Omega^{\omega}%
\in\Omega^{2}\left(  U_{\alpha},\text{ }Lie\;G\right)  .$

Under a change of local trivialization or gauge change, determined by the
transition function g$_{\alpha\beta}$ : $U_{\alpha}\cap U_{\beta}\rightarrow
G$, we obtain%
\[
A_{\beta}\text{ = }g_{\alpha\beta}^{-1}\text{ }A_{\alpha}\text{ }%
g_{\alpha\beta}\text{ + }g_{\alpha\beta}^{-1}\text{ }dg_{\alpha\beta}\text{ ,
}F_{\beta}\text{ = }g_{\alpha\beta}^{-1}\text{ }F_{\alpha}\text{ }%
g_{\alpha\beta}%
\]
...and if a cover by local trivialization and a set of potentials and fields
satisfying the above relations are given, we can reconstruct the connection
and the curvature in the principal fiber bundle, furthermore, if the cover and
the set of transition function are given, we can rebuild the total space of
the fiber bundle.

The eletromagnetic theory is a special case of gauge theory where $G=U(1)$,
this inner degree of freedom being interpreted by the helicity of the photon
since Weinberg \cite{13}, in such a way that a choice of a set of gauge
potentials in M could be interpreted by the changing from an empty Minkowsky
space by one with plenty of photons.

In the general case, the homogeneous field equations, analogous to the
Maxwell's homogeneous equations are given by the pull-back, by local sections,
of the Bianchi's identity, $D^{\omega}\Omega^{\omega}=0$, consequence of
Jacobi's identity and Poincar\'{e} lemma. These are simply the consequence
from the assumption that the field can be derived from a potential.

We now define in the total space of the fiber bundle the natural metric%
\[
h\left(  X\text{, }Y\right)  \text{ = }\pi^{\ast}g\left(  X\text{, }Y\right)
\text{ }+k\text{ }\left(  \omega\left(  X\right)  ,\omega\left(  Y\right)
\right)  \text{ }(1)
\]
where g is the metric in the base space M and k is the bi-invariant Killing
form on G. Since this last one is negative definite, the total space has
signature $(1,\;3+dim\;G)$ as pointed above.

From the first part of the projection theorem as in page 144 of Bleecker [2]
we have, for each geodesic curve $s\rightarrow\gamma(s)$ in the total space of
the bundle with the metric defined in (1), that $\omega(\gamma^{\prime}(s))$ =
$Q\in Lie\;G$ is a constant momentum in the fiber direction. This is
interpreted as the Lie valued gauge charge of the particle.

From the second part of the projection theorem, this geodesic projects onto a
gauge charged particle's motion $x(s)=\pi\circ\gamma(s)$\bigskip, for which
the velocity $v=dx/ds$ obeys%
\[
\frac{Dv}{ds}\text{ = }[F(x(s))]\text{ }v(s)\text{ }(2)
\]
where the 4x4 matrix field $F$ is given in each gauge choice by
$F=k(Q,\;\Omega_{\alpha}^{\omega})$. This corresponds to the Lorentz force in
the electromagnetic field, and applied to the velocity, gives the Levi-Civita
covariant derivative of such velocity. From the transformation rules and from
the invariance properties of the Killing form, $F$ does not depend on the
gauge choice and we can write $F=k(Q,\;\Omega^{\omega}$).

From the proof of the projection theorem, for each motion of a charged
particle in the base manifold and a given $Lie\;G$ valued charge $Q$ we get a
unique geodesic in the total space.

In what follows we consider the Goebel-Zeeman topology in the total space of
the fiber bundle and from it obtain a topology in the base manifold for wich
the continuous curves correspond to motions of charged particles in the base manifold.

\section{\textbf{Zeeman topology for Gauge Theories}}

Now we consider the fiber bundle structure of a Gauge Theory with a given
connection as in part 2 and the Zeeman-Goebel topology for the total space of
the fiber bundle. Therefore the continuous curves $\gamma:[a,b]\rightarrow P$,
are geodesic polygonals, where geodesic means geodesic with respect to the
metric given by $(1)$.

Since we have that $M=P/G$, from the Zeeman-Goebel topology in P and the usual
Group topology for $G$, we define the quotient topology in $M$. This is the
finest topology in $M$ that makes the projection $\pi:P\rightarrow M$
continuous. With this topology the only continuous curves are the paths of
charged particles.

For this let $x(s)$ the motion of a particle with gauge charge Q. We take the
unique geodesic lift of that curve satisfying $\omega$($\gamma$'$(s))=Q$,
since it is geodesic, $\gamma(s)$ and therefore $x(s)$ is continuous.

If $x(s)$ is a continuous curve in $M$, then for a local section
$\sigma_{\alpha}$, $\sigma_{\alpha}x$ is a continuous curve $\gamma(s)$ in the
principal bundle such that $x=\pi$ $\circ$ $\gamma$, and being $\gamma(s)$
continuous, it is a geodesic, therefore $x=\pi$ $\circ$ $\gamma$ corresponds
to the motion, under the given gauge field, of a particle with charge $\omega
$($\gamma$'$(s))=Q$. Note that if the particle is chargeless then the its path
is a continuous geodesic.

If $G=U(1)$, the electromagnetic case, the topology we have defined works as
well as Goebel topology for charged particles, being more naturally
constructed and more general than this. In what follows we employ the
following notation: $Z_{G}(P)$ is the Zeeman - Goebel topology on P (in which
continuous curves are geodesics) and $Z_{G}(M)$ is the projected Zeeman -
Goebel topology on M (in which the continuous curves are the paths of charged particles).

The Projected Zeeman Topology $Z_{\Pr}(M)$ is the strongest topology in which
the only continuous curves are paths of charged particles

Let $\tau^{\ast}(M)$ be a topology on $M$ in which the only continuous curves
are the paths of chaged particles. We can lift this topology to a topology on
$P$ \ ($\tau^{\ast}(P)$) as follows: Define $\tau^{\ast}(P)$ as the weakest
topology on $P$ such that the projection $\pi$ is continuous.

Let $\gamma$ denote a continuous curve on $(M,\tau^{\ast}(M))$ (and thus the
path of a chaged particle on $M$). Using the local trivialization we can lift
$\gamma$ to a continuous path $\gamma$' on $P$. We note that $\gamma$' is in
fact continuous in $\tau^{\ast}(P)$ since this topology is weaker then the
product topology on $UxG$. Now, by the projection Theorem $\gamma$' is a
geodesic in $P$ and it follows that geodesics on $P$ are continuous curves
(with topology $\tau^{\ast}(P)$). On the other hand, if $\gamma$' is a
continuos curve on $(P,\tau^{\ast}(P))$ then it projects to a continuos curve
on $(M,\tau^{\ast}(M)).$Thus, by the projection theorem, $\gamma$' is a
geodesic on $P$.

It follows that $\tau^{\ast}(P)<Z_{G}(P)$ since the Zeeman - Goebel topology
is the strongest topology in which the only continuous curves are geodesics.

This implies that $\tau^{\ast}(M)<Z_{\Pr}(M)$ since $\pi$ is an open map and
$Z_{\Pr}(M)$ is the strongest topology in which $\pi$ is continuous.

We conclude then that $Z_{G}(M)=$ $Z_{\Pr}(M)$.

\bigskip

\section{\textbf{Conclusion}}

We have presented a topology for which the continuous curves are polygonals of
motions of charged particles under gauge fields. If we define the physically
possible movements of any massive particle as $\ll$any continuous and
differentiable curve$\gg$ we obtain $\ll$the possible motions in gauge fields$\gg$ axiomatically from the topological structure of the space-time. This result
points in the direction of a completely geometric theory of particle motion.
It also suggest many directions to investigate, for example:

(1) on what is the relation between the group of homeomorphisms of this
topology and the group of gauge transformations,

(2) what would be the generalization and interpretation of typical gauge
theories ideas, as Berry's phases when the space-time has such topology,

(3) if it is possible to obtain classification theorems of fiber bundles with
a given structural group over a space-time with this topology, as it was done
for the instantons.

We note that in the case of instantons classification a more pragmatic than
physically acceptable employment of the four-dimensional sphere as a model for
space-time was done... but never objected.

\begin{acknowledgement}
We are grateful to Rold\~ao da Rocha Jr. for his revision and comments.
\end{acknowledgement}

\end{document}